\documentclass[twocolumn,prl,aps,superscriptaddress,showpacs]{revtex4}

\usepackage{bm}
\usepackage{graphicx}
\usepackage{ulem} 
\usepackage{color}
\def\urs{URu$_2$Si$_2$}

\def\prfep{PrFe$_4$P$_{12}$}

\def\updal{UPd$_2$Al$_3$}

\def\jpcm{J.\ Phys.\ Condens.\ Matter\ }
\def\jmmm{J.\ Magn.\ Magn.\ Mater.\ }
\def\ssc{Solid State Commun.\ }
\def\jpsj{J.\ Phys.\ Soc.\ Jpn.\ }
\def\pb{Physica B\ }

\def\nat{Nature\ }
\def\natp{Nature Physics\ }
\def\zpb{ZPhys. B\ }

\def\cond{cond-mat/}

\begin{document}

\title{A signature of hidden order in \urs\ : the excitation at the wavevector $\mathbf{Q_0} = (1\,0\,0)$}


\author{A. Villaume}
\affiliation{Institut Nanosciences et Cryogenie, SPSMS/MDN, 
CEA-Grenoble, 38054 Grenoble, France}

\author{F. Bourdarot}
\email{frederic.bourdarot@cea.fr}
\affiliation{Institut Nanosciences et Cryogenie, SPSMS/MDN, 
CEA-Grenoble, 38054 Grenoble, France}

\author{E. Hassinger}
\affiliation{Institut Nanosciences et Cryogenie, SPSMS/MDN, 
CEA-Grenoble, 38054 Grenoble, France}

\author{S. Raymond}
\affiliation{Institut Nanosciences et Cryogenie, SPSMS/MDN, 
CEA-Grenoble, 38054 Grenoble, France}

\author{V. Taufour}
\affiliation{Institut Nanosciences et Cryogenie, SPSMS/MDN, 
CEA-Grenoble, 38054 Grenoble, France}

\author{D. Aoki}
\affiliation{Institut Nanosciences et Cryogenie, SPSMS/MDN, 
CEA-Grenoble, 38054 Grenoble, France}

\author{J. Flouquet}
\affiliation{Institut Nanosciences et Cryogenie, SPSMS/MDN, 
CEA-Grenoble, 38054 Grenoble, France}

\date{\today}

\begin{abstract}

Simultaneous neutron-scattering and thermal expansion measurements on the heavy-fermion superconductor \urs\ under hydrostatic pressure of 0.67 \,GPa have been performed in order to detect the successive paramagnetic, hidden order, and large moment antiferromagnetic phases on cooling. The temperature dependence of the sharp low energy excitation at the wavevector $\mathbf{Q_0}=(1\,0\,0)$ shows that this excitation is clearly a signature of the hidden order state. In the antiferromagnetic phase, this collective mode disappears. The higher energy excitation at the incommensurate wavevector $\mathbf{Q_1}=(1.4\,0\,0)$ persists in the antiferromagnetic phase but increases in energy. The collapse of the inelastic neutron scattering at $\mathbf{Q_0}$ coincides with the previous observation of the disappearance of superconductivity. 


\end{abstract}


\maketitle

The elucidation of the nature of a hidden order in exotic materials which belong often to the rich class of strongly correlated electronic systems is a hot subject as it can lead to the discovery of unexpected new order parameters. Debates exist on quite different proposals such as orbital hidden order in the heavy fermion system \urs\ \cite{chandra02}, multipolar ordering in rare earth skutterudites \cite{karam05} or "spin order accompanying loop current" in cuprates superconductors \cite{aji07}.


Due to the dual character of the $5f$ electrons in \urs\, between localized (leading to the possibility of multipolar ordering) and itinerant (possibility of large Fermi Surface instabilities), this compound has been the subject of a large variety of experiments\cite{amitsuka07}. At zero pressure, a phase transition occurs from the paramagnetic (PM) phase to a so called hidden order (HO) phase at a temperature $T_0\sim17.5$\,\,\,K. The hidden order label reflects the fact that this order may not be of dipolar origin. The order parameter is not yet determined: spin or charge density wave \cite{maple86, ikeda98, mineev05}, multipolar ordering \cite{santini94, ohkawa99, hanzawa07, kiss05}, orbital antiferromagnetism \cite{chandra02}, chiral spin state \cite{gorkov92}, and helicity order \cite{varma06} have been proposed. The long standing debate on the occurrence of a tiny ordered moment $M_0\sim0.02\,\mu_B$ per U atom at $T\to0$ \,K for the antiferromagnetic (AF) wavevector $\mathbf{Q_{AF}}=(0\,0\,1)$ seems to converge now towards an extrinsic origin directly related to the high sensitivity of \urs\ to pressure and stress (low critical pressure $P_x \sim 0.5$\,GPa) \cite{amitsuka07,matsuda01,takagi07,amato04}.

\begin{figure}
\includegraphics[width=76mm]{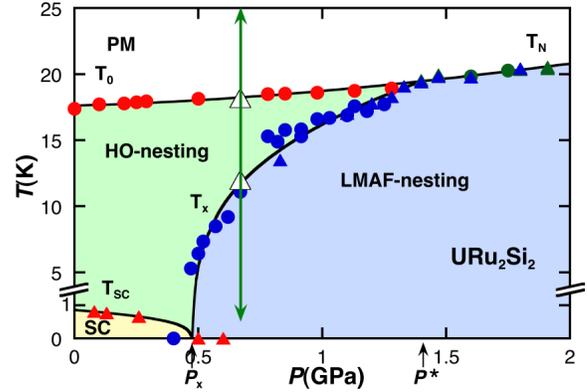}
\caption{(Color online) $(T, P)$ Phase diagram of \urs\ from resistivity (circles) and ac calorimetry (triangles) measurements \cite{hassinger08} with the low pressure HO phase and the high pressure AF phase. Bulk superconductivity state is suppressed at $P_x$ when antiferromagnetism appears. The open triangles correspond to the present determination of $T_0$ and $T_x$ at $P = 0.67$\,\,GPa.}
\label{phasedia}
\end{figure}

Pressure studies \cite{amitsuka07, motoyama03, bour05, hassinger08} reveal an interesting phase diagram (Fig.\ref{phasedia}). At $T\rightarrow0$\,K, neutron scattering experiments \cite{amitsuka07} show that the hidden order ground state switches at $P_x$ to a large moment antiferromagnetic (AF) state of sublattice magnetization $M_0$ near $0.3\,\mu_B$ with a propagation vector $\mathbf{Q_{AF}}$. The HO-AF boundary $T_x(P)$ meets the $T_0(P)$ line at the tricritical point ($T^\star\sim19.3\,\,K$, $P^\star\sim1.36$\,GPa) \cite{hassinger08}; above $P^\star$, a unique ordered phase (AF) is established under pressure below $T_N(P)$. Previous NMR experiments \cite{matsuda01,kohara86} as well as transport measurements \cite{maple86,hassinger08} indicate clearly that nesting occurs at $T_0$ as well as at $T_N$. 


The interest in \urs\ is reinforced by the appearance of unconventional superconductivity at $T_c \sim1.2$\,K for $P=0$ \cite{schlabitz86} which disappears in the bulk at $P_x$ \cite{hassinger08}.

Up to now, there is no direct convincing microscopic signature of the hidden order state. For example, the previous claim of residual Si NMR linewidth \cite{bernal01} has been rejected \cite{hassinger08b, takagi07}; the proposal of orbital antiferromagnetism is not demonstrated \cite{Wiebe04}.

The aim of the present work is to clarify the inelastic neutron scattering response for both ordered phases. At $P = 0$, two main inelastic magnetic responses of $\mathbf{Q_{AF}} = (0\,0\,1)$ and of an incommensurate wavevector $\mathbf{Q_{INC}} = (0.4\,0\,1)$ are detected. These signals are robust : rather insensitive to annealing conditions by contrast to the temperature dependence of the elastic intensity linked to the residual tiny ordered moment \cite{bour94} Because of the Ising character along the \textit{c} axis of the magnetic excitations, they has been measured at the equivalent positions $\mathbf{Q_0} = (1\,0\,0)$ for $\mathbf{Q_{AF}}$ and $\mathbf{Q_1} = (1.4\,0\,0)$ for $\mathbf{Q_{INC}}$ \cite{broholm87,broholm91, bour03, wiebe07}. The remarkable feature is that below $T_0$ both excitations are sharp with respective gaps at $\Delta_0=1.8$\,meV and $\Delta_1=4.5$\,meV \cite{bour03}. Furthermore their temperature evolutions explain the shape of the specific heat anomaly at $T_0$ \cite{wiebe07, vandijk97}. The clear trend is a strong interplay between these two inelastic responses. It was recently suggested that  $\mathbf{Q_{INC}}$ may be a wavevector for a spin density wave occurring at $T_0$ \cite{wiebe07}. However no evidence is found even in NMR experiments \cite{takagi07}.


Previous neutron scattering experiments under pressure have lead to suggest that the low energy excitation characteristic of $\mathbf{Q_0}$ may collapse at low temperature when entering into the antiferromagnetic state \cite{bour94,cond2003,amitsuka00}; but either the accuracy of the data is poor or the pressure condition is not well established. Futhermore there are contradictory conclusions for $\Delta_1$ : persistence according to \cite{cond2003} or collapse according to \cite{amitsuka00}. In contrast to these previous experiments \cite{amitsuka00,cond2003} where the studies are made at different pressures with no analysis of the temperature dependence, the present choice is to work at a constant pressure $P = 0.67$\,GPa slightly above $P_x$. At this pressure, each phase has a significant temperature range of existence as $T_0 = 18.2$\,K and $T_x = 12.0$\,K. Furthermore, the precise transition temperatures $T_0$ and $T_x$ have been determined during the neutron scattering experiment by thermal expansion. 


The main result of this letter is the simultaneous thermal evolution of the ordered antiferromagnetic moment and of the inelastic intensities of the gaps at $\mathbf{Q_0}$ and $\mathbf{Q_1}$ both in the hidden order and antiferromagnetic phase at $P=0.67$ \,GPa. The hidden order state is associated with a strong inelastic signal at $\mathbf{Q_0}$. In the antiferromagnetic phase, the inelastic signal vanishes. At $\mathbf{Q_1}$, a clear inelastic spectrum persists; the gap $\Delta_1$ changes abruptly when entering into the antiferromagnetic phase.

Neutron-scattering measurements were performed at the Institut Laue-Langevin (ILL) on the IN12 and IN22 cold and thermal triple-axis spectrometers, respectively. The energy resolution determined by the incoherent scan at elastic energy transfer was 0.22 meV and 0.9 meV respectively on IN12 and IN22 at full width at half maximum. Furthermore, as the IN12 spectrometer is located at the end of the cold-neutron-guided, the  background is far lower than the one in previous pressure experiments \cite{amitsuka07,metoki00}. IN22 was used for the studies at $\mathbf{Q_1}$ in order to reach higher energy. In each case, we used a well adjusted cadmium shielding around the pressure cell.

A single crystal, grown by the Czochralski method, of size of $\sim 5 \times 4 \times 3$ mm$^3$, from the same batch used in the high-field measurements of \cite{bour03} and in the previous high pressure measurements \cite{cond2003}, was used for the experiment. A flat surface was cleaved perpendicular to the $c$ axis. This axis and one $a$ axis were in the scattering plane.

Measurements under pressure were performed using a home-made CuBe pressure cell. A strain gage (see \cite{villaume07}) was glued along the $a$ axis on the flat surface perpendicular to the $c$ axis. The pressure dependent superconducting transition of lead, was measured by ac-magnetic susceptibility. To transmit the pressure, a mixture 1:1 of  fluorinert 70 \& 77 was used. The pressure conditions are similar to those used in reference \cite{amitsuka07}. 

\begin{figure}
\includegraphics[height=59mm]{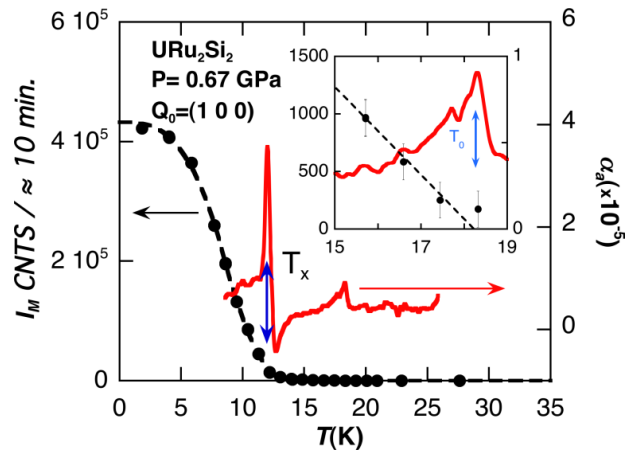}
\caption{(Color online) Thermal evolution of the magnetic Bragg peak intensity at $\mathbf{Q_0}=(1\,0\,0)$ (dashed line) and of the thermal expansion $\alpha_a$ (solid line) of \urs\ at $P = 0.67$ \,GPa. \textit{I$_M$} corresponds to the count at the top of the magnetic Bragg peak $\mathbf{Q_0}$ per 10 minutes, the background subtracted. The inset is a zoom of \textit{I$_M$} and $\alpha_a$ in the HO phase.}
\label{fig2}
\end{figure}

The thermal expansion measurements performed at  $P=0.67$ \,GPa indicate transition temperatures of $T_0 = 18.2$ \,K and $T_x = 12.0$ \,K (Figure \ref{fig2}). An excellent agreement is found between these results and the recent determination of the ($T, P$) phase diagram as shown in Fig.\ref{phasedia}\cite{hassinger08}. The precise knowledge of the localization in this phase diagram is an important advantage of this experiment to corroborate both, thermal expansion and neutron scattering experiments. This was not achieved in the previous experiments\cite{amitsuka00, cond2003}. The temperature dependence of the magnetic elastic intensity ($I_M \propto M^{2}$) at $\mathbf{Q_0}$ is also shown in Fig.\ref{fig2}. The onset of the large elastic antiferromagnetic signal at $\mathbf{Q_0}$ coincides with the thermal expansion jump at $T_x$. The estimation of $M_0$ in the antiferromagnetic phase gives  $0.4\,\mu_{B}/U$, in agreement with previous results. \cite{cond2003, bour04, bour05, amitsuka07}. A small magnetic intensity survives above $T_x$ and collapses linearly in temperature at $T_0$ as found in Ref. \cite{bour05, broholm87} but not in \cite{amitsuka07}. The extrapolation to 0\,K of the tiny ordered moment is  $0.05\,\mu_{B}/U$. This tiny ordered moment is assumed to emanate from the same extrinsic origin as the one at ambient pressure.


\begin{figure}
\includegraphics[height=58mm]{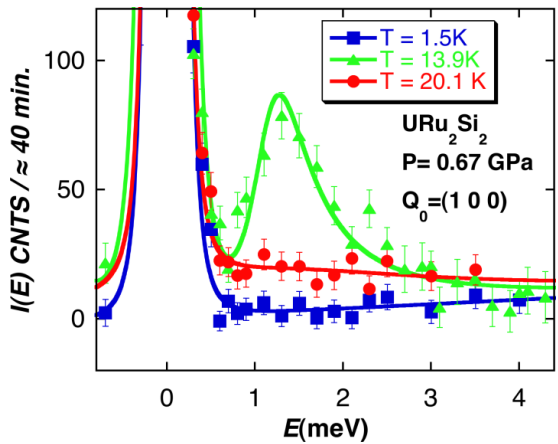}
\caption{(Color online) Energy scan at $\mathbf{Q_0}$ in the PM ($T=20.1$ \,K), HO ($T=13.9$ \,K), and AF ($T=1.5$ \,K) phases. Only electronic background has been subtracted. The curves are guides for the eyes.}
\label{scanin}
\end{figure}

A large inelastic signal, coming essentially from the sample as it can be verified on the residual background, is shown in Fig. \ref{scanin}. At $\mathbf{Q_0}$, in the paramagnetic regime ($T$=20.1\,K), the signal is weak and strongly damped. In the hidden order phase ($T$=13.9\,K), an inelastic spectrum similar to the spectrum measured in \urs\ at $P=0$ with an energy gap $E_0 \sim 1.25$ meV is observed. In the antiferromagnetic state ($T$=1.5\,K), neither quasi-elastic nor inelastic response can be detected at low temperature.

\begin{figure}
\includegraphics[height=58mm]{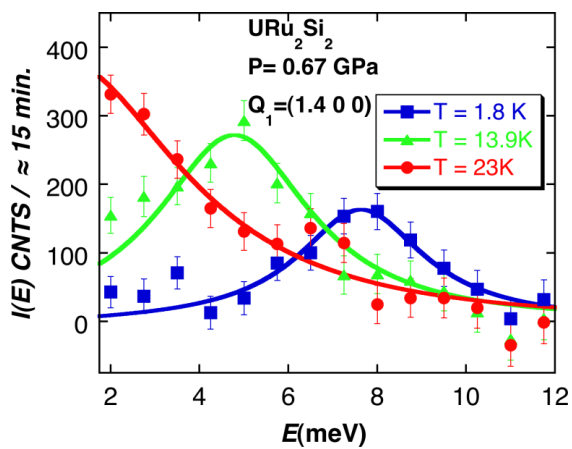}
\caption{(Color online) Energy scan at $\mathbf{Q_1}$ in the PM ($T=20.1$ \,K), HO ($T=13.9$ \,K), and AF ($T=1.5$ \,K) phases. An energy scan measured at $T = 1.8$ K at the wavevector $\mathbf{Q}=(1.3\,0\,0)$ has been used as background and subtracted to the energy scans measured at $\mathbf{Q_1}$. The curves are guides for the eyes.}
\label{Inela-Q1}
\end{figure}

Fig. \ref{Inela-Q1} represents the magnetic excitation at the wavevector $\mathbf{Q_1}$ in the three phases. Above $T_0$, the signal is mainly quasielastic and broadened. In the hidden order phase, it becomes mostly inelastic with an energy gap $\Delta_1\,=\,5$\,meV. This magnetic excitation persists on entering into the antiferromagnetic phase, but is shifted to 7.8\,meV. This shift to high energy corresponds to a decrease in the inelastic amplitude (at zero order, $\mathbf{I_{\Delta_1}}$ varies like $\frac{1}{\Delta_1}$ \cite{wiebe07})

\begin{figure}
\includegraphics[height=58mm]{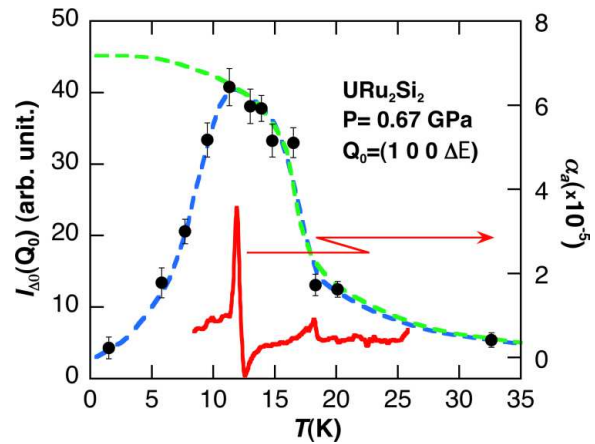}
\caption{(Color online) Integration of the dynamic suscetibility $I_{\Delta_0}(Q_0)$ from $E=0.6$ meV $\rightarrow E=2.5$ meV (black circles). The dashed blue curve is a guide for the eyes. Thermal expansion in red (solid line). If the HO persists down to 0 \,K, its contribution, measured at $P=0$ in reference \cite{bour94}, is the extrapolated green dotted line.}
\label{Iinela}
\end{figure}

In order to precisely characterize the inelastic response at $\mathbf{Q_0}$, we performed many scans in the temperature range 1.5\,K to 33\,K. Fig. \ref{Iinela} shows, $I_{\Delta_0}(\mathbf{Q_0}) \propto \int_{0.6 meV}^{2.5 meV}{\chi''(E,\mathbf{Q_0})}dE$,  the integration of the dynamic susceptibility at $\mathbf{Q_0}$ over the range of energy from 0.6 to 2.5 meV where the magnetic excitation is detected. In the paramagnetic state, the strongly damped signal increases smoothly on approaching  $T_0$. At $T_0$, the intensity rises abruptly. This increase is very similar to the behavior of the the integration of the dynamic susceptibility at $\mathbf{Q_0}$ found in the sample at $P=0$ \cite{bour05, broholm91, wiebe07}. At $P=0.67$\,GPa, the intensity reaches a maximum at $T_x$ and then decreases and collapses for $T \rightarrow 0$.

The increase of \textit{I$_M$} below $T_x$ down to 5 \,K is concomitant  with the decrease of $I_{\Delta_0}(Q_0)$. The smooth onset of the ordered antiferromagnetic moment at $T_0$ can have an intrinsic or extrinsic origin. By comparison with previous data \cite{amitsuka07,bour05}, it must be noticed than in \cite{amitsuka07}, the temperature variation of the elastic magnetic signal \textit{I$_M$} is slow just above $P_x$, rather steep near 0.8 \,GPa and again slow at 1 \,GPa. In \cite{bour05}, for the same pressure of 1 \,GPa, \textit{I$_M$} has a steeper $T$ dependence than in \cite{amitsuka07}. All the data point out a high sensitivity to the pressure inhomogeneity. Assuming that $I_M$ and $I_{\Delta_0}(Q_0)$ reflect respectively the AF fraction and the HO fraction, we can assert that at least $95\%$ of antiferromagnetic phase is achieved at 1.5 \,K. This evaluation confirms the conclusion of a recent report \cite{amitsuka08}.

Our result at $P=0.67$\,GPa is that the main feature of \urs\ occurs for both hidden order and antiferromagnetic phases at the  wavevector $\mathbf{Q_0}$ with sharp excitations in the hidden order phase and a large elastic magnetic signal in the antiferromagnetic state; the excitation collapses at low temperature in the antiferromagnetic state. It was proposed in the framework where the hidden order is quadrupolar \cite{ ohkawa99} that the strong excitations describe the longitudinal fluctuations of a magnetic dipole, whereas, inside the antiferromagnetic phase, the inelastic neutron scattering signal coming from quadrupolar fluctuations is not measurable by neutron scattering.

The observation of the excitation $\Delta_1$ at $\mathbf{Q_1}$ even in the antiferromagnetic phase appears correlated with the persistence of nesting through $P_x$ derived from transport measurements. Without nesting, the system will end up in the paramagnetic ground state. The lost of electronic carriers at $T_0$ or $T_N$ changes the damped response at $\mathbf{Q_0}$ and $\mathbf{Q_1}$ in the paramagnetic state to well defined excitations below the onset of the long range ordering. A nesting is an associated necessary condition for the restoration of the local properties of the 5f electrons of the U atoms. $\mathbf{Q_0}$ seems to be the ordered wavevector for both hidden order and antiferromagnetic phases. In the case of a switch from a spin density wave state at $\mathbf{Q_1}$ to an antiferromagnetic state at $\mathbf{Q_0}$ above $P_x$, it is expected to be accompanied by a drastic change of the excitations at $\mathbf{Q_1}$. 




It is worthwhile to compare \urs, where the exotic properties originate from the $5f^2$ configuration of the U atoms, with new Pr skutterudites systems, where now the key electrons belong to the $4f^2$ configuration. The situation of \urs\ at $T_0$ seems to be similar to that reported for \prfep\ \cite{hassinger08b,park05,kohgi06} as regards the concomitant effect of nesting and HO parameter at the ordering temperature $T_A$. Furthermore, it is interesting to mention that in \prfep\, it has been established that a switch from HO to antiferromagnetic state occurs at $P_x \sim 2$\,GPa, the key wavevector being  $\mathbf{Q}=(1\,0\,0)$ for both HO and AF phases \cite{hidaka05}.

Knowing the result that bulk superconductivity disappears also at $P_x$ as demonstrated by microcalorimetry experiment \cite{hassinger08}, it is appealing to claim that the low energy excitation at $\mathbf{Q_0}$ is the origin of the superconductivity : The simple idea is that, as proposed for \updal\ \cite{miyake01,sato01,chang07}, the low energy mode plays the role of an exciton mode with favorable Cooper pairing potential. 


To conclude, our experiments point out the drastic change in the response of the inelastic neutron scattering at $\mathbf{Q_0}$ precisely at the transition from hidden order to antiferromagnetism. The collective low energy mode at $\mathbf{Q_0}$ is a signature of the hidden order phase. 
The next step will be to analyze among all possible order parameters for hidden order, which one can lead to the present excitations : collapse for $Q_0$ and persistence for $Q_1$ when entering the antiferromagnetic state. Another issue will be to precise the band structure in the different phases depending of the hypothesis on the localization of the $5f$ electrons in order to clarify the Fermi Surface nesting. Finally, let us emphasize the great advantage of simultaneous thermal expansion and neutron scattering experiments. Our technique can now be applied to other relevant examples like tiny ordered moments at a quantum critical point.

This work is supported by the french research through ANR-06-BLAN-0220.

\end{document}